\documentclass[twocolumn]{aastex63}

\usepackage{amsmath,amsbsy}
\usepackage{url} 
\usepackage{natbib,hyperref}
\usepackage{diagbox}

\received{\today}

\shorttitle{Curvature \textit{MMS}}
\shortauthors{Bandyopadhyay et al.}

\begin{document}

\title{In situ Measurement of Curvature of Magnetic Field in Turbulent Space Plasmas: A Statistical Study}

\author[0000-0002-6962-0959]{Riddhi Bandyopadhyay}
\email{riddhib@udel.edu}
\affiliation{Department of Physics and Astronomy, University of Delaware, Newark, DE 19716, USA}
\affiliation{Bartol Research Institute, University of Delaware, Newark, DE 19716, USA}	

\author[0000-0002-6962-0959]{Yan Yang}
\affiliation{Southern University of Science and Technology, Shenzhen, Guangdong 518055, China}
\affiliation{University of Science and Technology of China, Hefei, Anhui 230026, China}

\author[0000-0001-7224-6024]{William~H. Matthaeus}
\affiliation{Department of Physics and Astronomy, University of Delaware, Newark, DE 19716, USA}
\affiliation{Bartol Research Institute, University of Delaware, Newark, DE 19716, USA}	

\author[0000-0001-8478-5797]{Alexandros Chasapis}
\affiliation{Laboratory for Atmospheric and Space Physics, University of Colorado Boulder, Boulder, Colorado, USA}

\author[0000-0003-0602-8381]{Tulasi N. Parashar}
\affiliation{Department of Physics and Astronomy, University of Delaware, Newark, DE 19716, USA}
\affiliation{Bartol Research Institute, University of Delaware, Newark, DE 19716, USA}	

\author[0000-0003-1639-8298]{Christopher T. Russell}
\affiliation{University of California, Los Angeles, California 90095-1567, USA}

\author[0000-0001-9839-1828]{Robert J. Strangeway}
\affiliation{University of California, Los Angeles, California 90095-1567, USA}

\author[0000-0001-7188-8690]{Roy B. Torbert}
\affiliation{University of New Hampshire, Durham, New Hampshire 03824, USA}

\author[0000-0001-8054-825X]{Barbara L. Giles}
\affiliation{NASA Goddard Space Flight Center, Greenbelt, Maryland 20771, USA}

\author[0000-0003-1304-4769]{Daniel J. Gershman}
\affiliation{NASA Goddard Space Flight Center, Greenbelt, Maryland 20771, USA}

\author[0000-0001-9228-6605]{Craig J. Pollock}
\affiliation{Denali Scientific, Fairbanks, Alaska 99709, USA}

\author[0000-0002-3150-1137]{Thomas E. Moore}
\affiliation{NASA Goddard Space Flight Center, Greenbelt, Maryland 20771, USA}

\author[0000-0003-0452-8403]{James L. Burch}
\affiliation{Southwest Research Institute, San Antonio, Texas 78238-5166, USA}




\begin{abstract}
Using in situ data, accumulated in the turbulent magnetosheath
by the \textit{Magnetospheric Multiscale (MMS)} Mission, we report a statistical study of magnetic field curvature and discuss its role in the turbulent space plasmas. Consistent with previous simulation results, the Probability Distribution Function (PDF) of the  curvature is shown to have distinct power-law tails for both high and low value limits. We find that the magnetic-field-line curvature is intermittently distributed in space. High curvature values reside near weak magnetic-field regions, while low curvature values are correlated with small magnitude of the force acting normal to the field lines. A simple statistical treatment provides an explanation for the observed curvature distribution.  This novel statistical characterization of 
magnetic curvature in space plasma provides a starting point for assessing, in a turbulence
context,
the applicability and impact of particle energization processes, such as curvature drift, that rely on this fundamental quantity.
\end{abstract}



\section{Introduction} \label{sec:intro}
The curvature of the magnetic field
enters in numerous important ways in electrodynamics~\citep{Petschek1964} and plasma physics~\citep{Boozer2005RMP}, representing one of
the principle ways that magnetic fields
interact with matter. 
Curvature plays a key role in 
magnetic reconnection~\citep{Petschek1964},
stability of magnetic confinement~\citep{Dobrott1977PRL},
in magnetospheric physics and space physics~\citep{Hameiri1991JGR},
and in particle 
heating and acceleration~\citep{Jokipii1982ApJ, Pesses1981ApJL, Dahlin2014PoP}.
Usually, curvature is studied with regard to specific 
magnetic configurations.
For example, 
stability with respect to 
ballooning modes requires favorable
curvature that is antiparallel to
the pressure gradients~\citep{Boozer2005RMP}.
Similarly, the large curvature of field
lines in reconnection exhausts
gives rise to relaxation towards a
less stressed state, leading to 
electron energization by curvature drift acceleration~\citep{Dahlin2014PoP}. Magnetic-field curvature has been useful for detecting helical field configuration of flux ropes from in situ measurements~ \citep{Slavin2003JGR, Shen2007JGR, Sun2019GRL}. 

Recently, the curvature of magnetic field lines has been studied in 
the magnetohydrodynamic (MHD) model of plasma
turbulence~\citep{Yang2019ArXiv}. 
In the case of turbulence, it is impractical to 
study curvature of individual field 
lines and one may resort to 
a statistical approach, as is
typical in studies of turbulence~\citep{MoninYaglom-vol1}.
In these simulations, 
one finds interesting properties
such as a 
distribution
of curvature that 
exhibits two power-law regimes,
and a systematic anticorrelation of curvature with magnetic
field strength, for low values of magnetic field strength.
Here, we extend this statistical examination of magnetic 
curvature by analysis 
of {\it in situ} 
satellite observations in the 
terrestrial magnetosheath.
We employ \textit{Magnetospheric Multiscale (MMS)} data 
that reveal distributions and correlations
that are consistent with, and in fact very similar to,
those observed in the MHD simulations~\citep{Yang2019ArXiv}.
These results confirm the
theoretical model given in \cite{Yang2019ArXiv},
opening the door to new
applications
such as 
curvature drift acceleration
in turbulence as well
as the possible role of local 
explosive instabilities in turbulence.

The outline of the paper is as follows: in Sec.~\ref{sec:theory}, we discuss the theoretical derivation and approximations. In Sec.~\ref{sec:MMS}, we apply the theoretical constructs to MMS data and present the results. We discuss the importance of the results and conclude in Sec.~\ref{sec:disc}. In Appendix~\ref{sec:fndist}, \ref{sec:gaussian}, we justify the assumptions made in Sec.~\ref{sec:theory}. Appendix~\ref{sec:add}  shows the results presented in Sec~\ref{sec:MMS} for a different MMS interval.

\section{Theory and Method} \label{sec:theory}
The curvature $\kappa$ of magnetic field $\mathbf{B}$ is defined as
\begin{eqnarray}
\kappa = |\mathbf{b} \cdot \mathbf{\nabla} \mathbf{b}| \label{eq:curv},
\end{eqnarray}
where $\mathbf{b} = \mathbf{B}/B$ and $B=|\mathbf{B}|$.
It can be expressed also in the form
\begin{eqnarray}
\kappa &=& \frac{|\mathbf{b} \times (\mathbf{B} \cdot \mathbf{\nabla} \mathbf{B})|}{B^2} \nonumber
\\
&=& \frac{f_n}{B^2} \label{eq:fn},
\end{eqnarray}
where $f_n = |\mathbf{b} \times (\mathbf{B} \cdot \mathbf{\nabla} \mathbf{B})|$ is the magnitude of the  
tension force (per unit volume)
acting normal to the field lines.
In the curvilinear coordinate attached to a field line, traced
by a trajectory $\mathbf{\gamma}(s)$, the scalar 
$s$ is a coordinate along the field line, while
$\mathbf{e}_t = \frac{d{\bf \gamma}}{ds} / |\frac{d{\bf \gamma}}{ds}| = \mathbf{b}$ 
and  $\mathbf{e}_n = \frac{d^2{\bf \gamma}}{ds^2} / |\frac{d^2{\bf \gamma}}{ds^2}|$ 
are the unit vectors in the tangential and normal directions along the field
line, respectively. Then
\begin{eqnarray}
\mathbf{B} \cdot \mathbf{\nabla} \mathbf{B} = 
(B \mathbf{e}_t) \cdot \mathbf{\nabla} (B \mathbf{e}_t) = 
B \frac{\partial (B \mathbf{e}_t)}{\partial s} = 
B \frac{\partial B}{\partial s} \mathbf{e}_t - \kappa B^2 \mathbf{e}_n.
\label{eq:Bgrad}
\end{eqnarray}
Equation (\ref{eq:fn}) follows directly 
from Eq.~(\ref{eq:Bgrad}).

It is shown in the following section that high curvature values are well associated with weak magnetic field.
In contrast, 
low curvature values mainly result from small normal force, 
more so than from large values of magnetic field.
These findings point the way to explain the power-law tails in the 
curvature distribution in 
both the high value range, and the low 
value range, reasoning as follows \cite{Yang2019ArXiv}. 

First, let us consider
the low value range.
Noting that 
the normal force is two-dimensional,
we write its cartesian components as $f_1$ and $f_2$, and then assume that these are independent random variables and their PDFs for small values obey Gaussian distribution. Then the PDFs of $f_1$ and $f_2$ at small values may be written as
\begin{eqnarray}
P_{f_1}(f)=P_{f_2}(f)=\frac{1}{\sqrt{2\pi \sigma_1^2}} e^{-\frac{f^2}{2\sigma_1^2}} \label{eq:Pf12},
\end{eqnarray}
where, $f$ denotes the vale of either $f_1$ or $f_2$ at the point of interest, and $\sigma_1^2$ is the variance.
The quantity $f_n^2/\sigma_1^2=(f_1^2+f_2^2)/\sigma_1^2$ should then be distributed according to the chi-squared
distribution with 2 degrees of freedom 
The corresponding PDF of $f_n$ at small values (i.e. $f_n \rightarrow 0$) is
\begin{eqnarray}
P_{f_n}(f)=\frac{f}{\sigma_1^2} e^{-\frac{f^2}{2\sigma_1^2}} \label{eq:Pf}.
\label{eq:pdf-fn}
\end{eqnarray}
Here, $f$ represents the value of the variable $f_n$. 
See Appendix~\ref{sec:fndist} where a slightly 
more general, but equivalent, development is given.
Since $\kappa=f_n B^{-2}$ and low curvature $\kappa$ is determined by the scaling behavior of small normal force $f_n$,
the PDF of curvature as $\kappa \rightarrow 0$ can be written as
\begin{eqnarray}
P_{\kappa}(\kappa^{\prime})=B^2 P_{f_n}(\kappa^{\prime} B^2)=\frac{B^4 \kappa^{\prime}}{\sigma_1^2} e^{-\frac{B^4 {\kappa^{\prime}}^2}{2\sigma_1^2}}.
\label{eq:pdf-low-k}
\end{eqnarray}
Here, $\kappa^{\prime}$ is the value of the variable $\kappa$. Let us 
assume that $B$ is finite in this limit, which could be replaced with $B_{\mathrm{rms}}$ in equation (\ref{eq:pdf-low-k}).
Then the
Taylor series of the PDF around $\kappa^{\prime}=0$ is
\begin{eqnarray}
\frac{B^4}{\sigma_1^2} \left(\kappa^{\prime} -\frac{B^4}{2\sigma_1^2} {\kappa^{\prime}}^3 + \cdots \right) \label{eq:lowk-taylor}.
\end{eqnarray}
The higher-degree terms are 
much
smaller as $\kappa^{\prime} \rightarrow 0$, so we retain only the lowest order term,
and obtain
\begin{eqnarray}
P_{\kappa \rightarrow 0}(\kappa^{\prime}) \sim {\kappa^{\prime}}^1 \label{eq:Pklow}.
\end{eqnarray}

In a similar way, we can explain the power-law tail of the PDF for high curvature values.
In isotropic turbulence, we suppose that 
$x$, $y$, and $z$ components of magnetic fluctuations are 
independent Gaussian random variables.
Then
\begin{equation}
B_x,~B_y,~B_z \sim \mathcal{N}(0, \sigma_2^2) \label{eq:Bxyz},
\end{equation}
where $\mathcal{N}(0, \sigma_2^2)$ denotes the normal distribution with mean $0$ and variance $\sigma_2^2$. In real systems, the magnetic fields are never fully isotropic, so we eliminate the average from each component and work with the fluctuations. Note that, in a turbulent system, the \textit{increments} of the  magnetic field are intermittently distributed with super-Gaussian tails~\citep{Matthaeus2015PTRSA}, but the fluctuation components themselves are rather well described by Gaussian distribution~\citep[][also see Appendix \ref{sec:gaussian}]{Batchelor51,PadhyeEA01}.
The quantity $B^2/\sigma_2^2=(B_x^2+B_y^2+B_z^2)/\sigma_2^2$, therefore, follows the chi-squared distribution with 3 degrees of freedom.
The corresponding PDF of $B^2$ is 
\begin{eqnarray}
P_{B^2}(b^{\prime})=\frac{\sqrt{b^{\prime}}}{(2\sigma_2^2)^{3/2}~ \Gamma(3/2)} e^{-\frac{b^{\prime}}{2\sigma_2^2}} \label{eq:Pbsqd}, 
\end{eqnarray}
where, $b^{\prime}$ represents the value of $B^2$ at the point of interest, and $\Gamma$ is the gamma function. 
Since $\kappa=f_n B^{-2}$ and 
high curvature $\kappa$ is determined by the scaling behavior of weak magnetic field $B^2$,
the PDF of curvature as $\kappa \rightarrow \infty$ can be written as
\begin{eqnarray}
P_{\kappa}(\kappa^{\prime})=\frac{f_n}{{\kappa^{\prime}}^2} P_{B^2}\left(\frac{f_n}{\kappa^{\prime}}\right) = \frac{f_n^{3/2}~ {\kappa^{\prime}}^{-5/2}}{(2\sigma_2^2)^{3/2}~ \Gamma(3/2)} e^{-\frac{f_n}{2\sigma_2^2}}.
\label{eq:pdf-high-k}
\end{eqnarray}
Again, $\kappa^{\prime}$ is the value of the variable $\kappa$.
In analogy to the prior case,
we 
assume that $f_n$ remains finite in this limit, 
and replace the associated  
value with the 
average $\langle f_n \rangle$ in equation (\ref{eq:pdf-high-k}).
Then the Taylor series for the 
PDF about $1/{\kappa^{\prime}}=0$ becomes, 
\begin{eqnarray}
\frac{f_n^{3/2}}{(2\sigma_2^2)^{3/2}~ \Gamma(3/2)} \left( {\kappa^{\prime}}^{-5/2} - \frac{f_n}{2\sigma_2^2} {\kappa^{\prime}}^{-7/2} + \cdots \right) \label{eq:highk-taylor}.
\end{eqnarray}
It follows that 
in the limit
as $\kappa^{\prime} \rightarrow \infty$, i.e. $1/{\kappa^{\prime}} \rightarrow 0$, the curvature PDF scales as
\begin{eqnarray}
P_{\kappa \rightarrow \infty}(\kappa^{\prime}) \sim {\kappa^{\prime}}^{-5/2} \label{eq:Pk-high}.
\end{eqnarray}
Previously~\citep{Yang2019ArXiv},
the above reasoning was found to 
explain the behavior of the 
distributions of curvature in three dimensional, isotropic, MHD simulations.
We now extend this inquiry to the case of a 
naturally occurring space plasma, the turbulent magnetosheath.

Below, we use four-spacecraft linear estimates of gradient, 
similar to the ``curlometer" method \citep{Dunlop1988ASR,Paschmann1998ISSI} to calculate $\mathbf{\nabla} \mathbf{b}$. 
Then a dot product with $\mathbf{b}$ yields the curvature $\kappa = |\mathbf{b} \cdot \mathbf{\nabla} \mathbf{b}|$. In section~\ref{sec:MMS}, we use this approach to 
analyze the statistical properties of 
the curvature field using \textit{MMS} observations, including the accuracy of the above scaling arguments. 

\section{\textit{MMS} Observations} \label{sec:MMS}
\textit{MMS} consists of four identical spacecraft orbiting the Earth, for the 
chosen period,
in a tetrahedral formation with small ($\sim$10 km) separation. 
The four \textit{MMS} spacecraft sample the near-Earth plasma including the magnetosheath~\citep{Burch2016SSR}. The Fast Plasma Investigation (FPI)~\citep{Pollock2016SSR} instrument calculates the proton and electron three dimensional velocity distribution functions (VDF) and the Flux-Gate Magnetometer (FGM)~\citep{Russell2016SSR} measures the vector magnetic field.

In burst mode, the Dual Ion Spectrometer (DIS) and the Dual Electron Spectrometer (DES) in FPI/\textit{MMS} measure the ion and electron VDF at cadence of $150~\mathrm{ms}$ and $30~\mathrm{ms}$, respectively. Plasma moments are calculated from each VDF at the 
corresponding time resolution. 
The time resolution of the FGM magnetic field is $128~\mathrm{Hz}$ in burst mode. 

\begin{figure*} 
	\begin{center}
		\includegraphics[width=\textwidth]{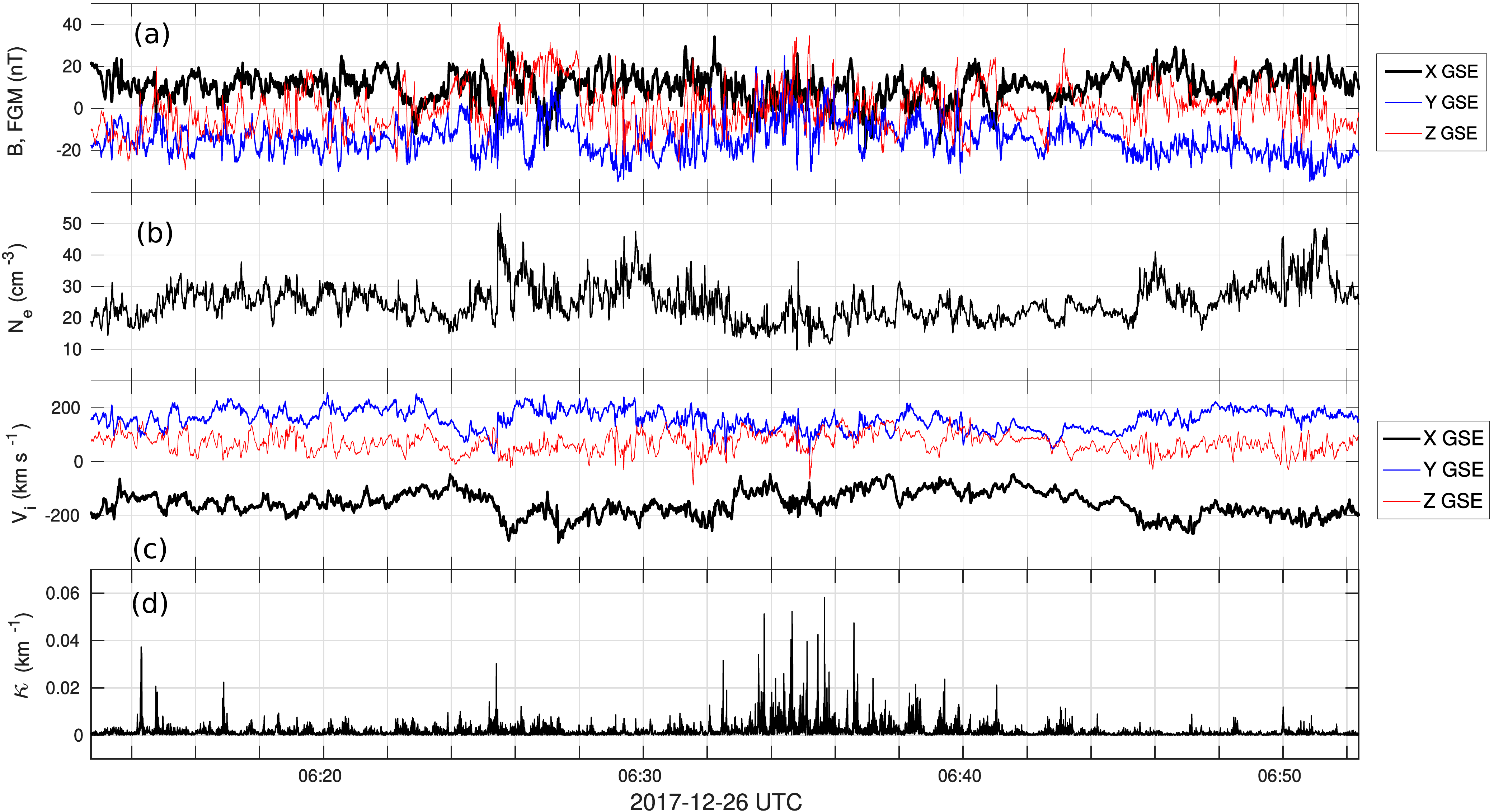}
		\caption{Overview of the \textit{MMS} observations in turbulent magnetosheath selected for this study. The data shown are from the FGM and FPI instruments on-board the \textit{MMS}1 spacecraft. Panel (a) shows the magnetic field measurements in GSE coordinates. Panel (b) shows the electron density. Panel (c) shows the ion velocity in GSE coordinates. Panel (d) shows the curvature calculated (approximately using a 
			curlometer-like method)
			from the magnetic field (see equation~(\ref{eq:curv}) ).}
		\label{fig:overview}
	\end{center}
\end{figure*}

To cover a large statistical sample of the turbulent plasma in the magnetosheath, here 
we focus on one 
long \textit{MMS} burst-mode interval between 06:12:43 and 06:52:23 UTC on 26 December 2017. A time series plot of the selected interval is shown in figure~\ref{fig:overview}. The FGM magnetic field components, in geocentric solar ecliptic (GSE) coordinate system~\citep{Franz2002PSS}, are shown in panel (a). The magnetic field components exhibit large-amplitude fluctuations which is typical for magnetosheath plasma. The electron density estimates are often more accurate than the ion density in the magnetosheath due to higher thermal speed. Panel (b) plots the electron density, obtained from the FPI/DES distributions. The three GSE components of the ion velocity components, measured by FPI/DIS, are plotted in panel (c). The final panel (d), shows the time series of the curvature field, derived from the magnetic field by a finite difference curlometer-like method (see Section~\ref{sec:theory}). The curvature values are observed to be 
highly intermittent with thin ``spikes" distributed during the whole interval that 
suggest the presence of sheet-like structures.

\begin{table}
	\caption{Description of \textit{MMS} Dataset from 06:12:43 to 06:52:23 UTC on 26 December 2017. The quantities are defined in the text.} 
	\centering
	\label{tab:overview}
	\begin{tabular}{c c c c c c}
		\hline 
		 MMS position & $|\langle \mathbf{B} \rangle|$ & $B_{\mathrm{rms}} /|\langle \mathbf{B} \rangle|$
		& $L$  
		& $d_{\mathrm{i}}$ 
		& $\beta_{\mathrm{p}}$ \\
	    (X, Y)$_{\mathrm{GSE}}$ & ($\rm nT$) &  &  $(\mathrm{km})$ & ($\rm km$)  &  \\
		\hline
		 $(10\, \mathrm{R_e}, 9\, \mathrm{R_e})$ & $17.9$ & $0.8$ & $27$ & $47$ & $4.4$\\
		\hline
	\end{tabular}
	\label{table1}
\end{table}

Several important plasma parameters of the selected \textit{MMS} interval are 
reported in Table~\ref{tab:overview}, including the locations of the MMS spacecraft in GSE coordinate system, in units of Earth radius $(\mathrm{R_{E}})$ during the interval, mean magnetic 
field $(|\langle \mathbf{B} \rangle|)$, ratio of rms fluctuation 
amplitude of the magnetic field $(B_{\mathrm{rms}})$ to the mean magnetic 
field, average spacecraft separation $L$, ion-inertial length $(d_{\mathrm{i}})$, and the average plasma beta $(\beta_{\mathrm{p}})$. The rms fluctuation amplitude is defined as $B_{\mathrm{rms}} = \sqrt{\langle |\mathbf{B}(t) - \langle \mathbf{B} \rangle|^2 \rangle}$, which has a value of $14\,\mathrm{nT}$ here. 
The density fluctuation amplitude is similarly defined. 
Note  that the density fluctuations are rather low, 
comparable to the typical values observed in the interplanetary solar wind. The spacecraft separations are much smaller than the ion-inertial length. However, using the curlometer method in the determination of gradients, there are some errors associated, which affect the estimation of $f_n$ and $\kappa$. Following \citep{Shen2003JGR} and \cite{Shen2007JGR}, the fractional error in the curvature values can be estimated roughly as,
\begin{eqnarray}
\frac{\Delta \kappa}{\kappa} \sim \mathcal{O}(L \kappa)\label{eq:kerr},
\end{eqnarray}
where $L$ is the spacecraft separation. For the present interval, if we use $L \sim 10\,\mathrm{km}$, and from Fig.~\ref{fig:overview}, the maximum of curvature values reaches about $\kappa \sim 0.5\,\mathrm{km}^{-1}$. Therefore, the fractional error in curvature remains within $\sim 0.5$. Further, by comparing the FPI current and curlometer current, several studies have found that MMS curlometer usually works well in the magnetosheath~\citep[e.g.,][]{Gershman2018PoP, Stawarz2019ApJL}. Therefore, the results presented below are expected to be reliable.

\begin{figure}
	\begin{center}
		\includegraphics[width=\linewidth]{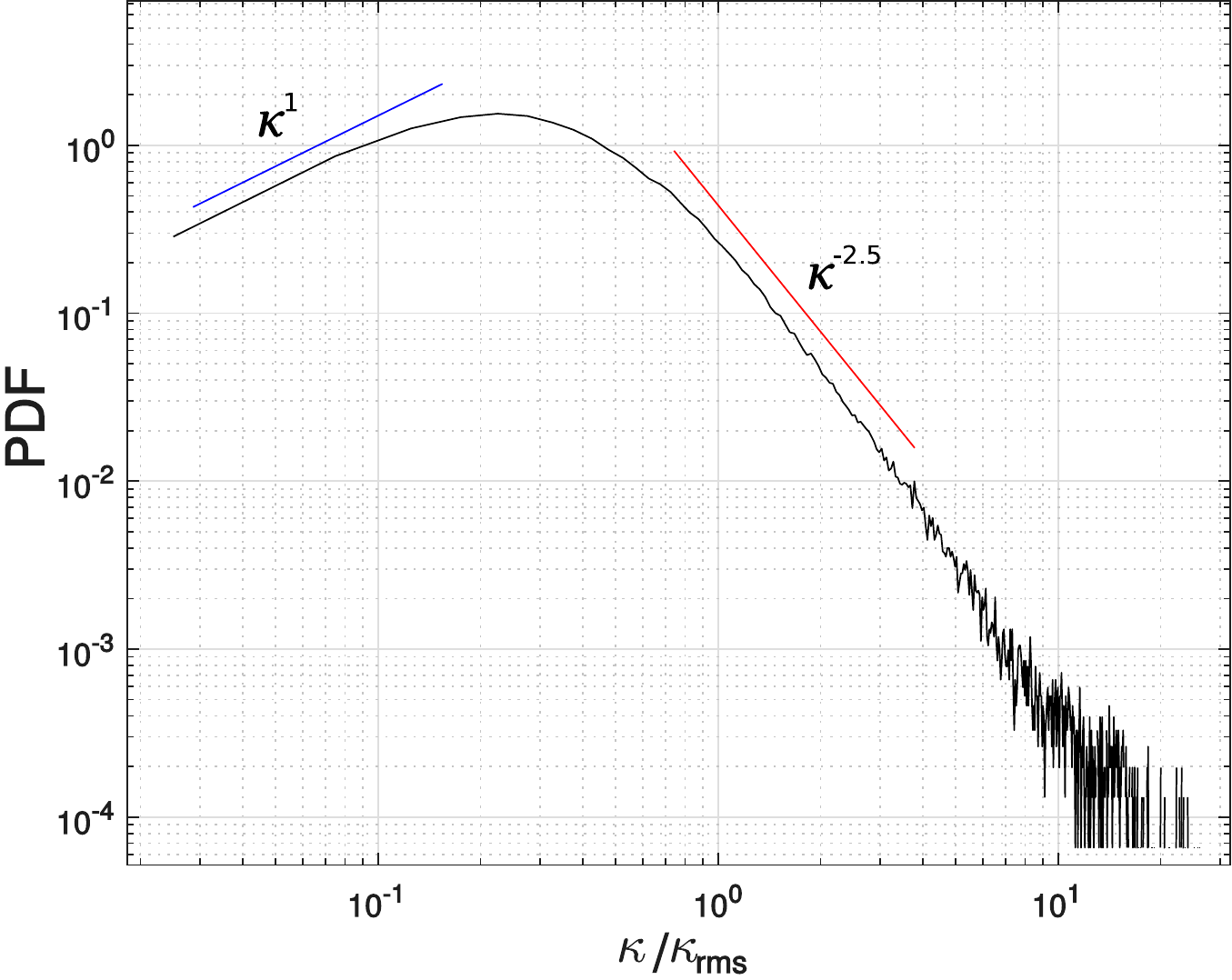}
		\caption{PDF of the magnetic field curvature $\kappa$ normalized to its rms value $\kappa_{\mathrm{rms}}$, computed from the 40 minute MMS dataset shown in Fig. \ref{fig:overview}.}
		\label{fig:curv_pdf}
	\end{center}
\end{figure}

\begin{figure*}
	\begin{center}
		\includegraphics[width=\linewidth]{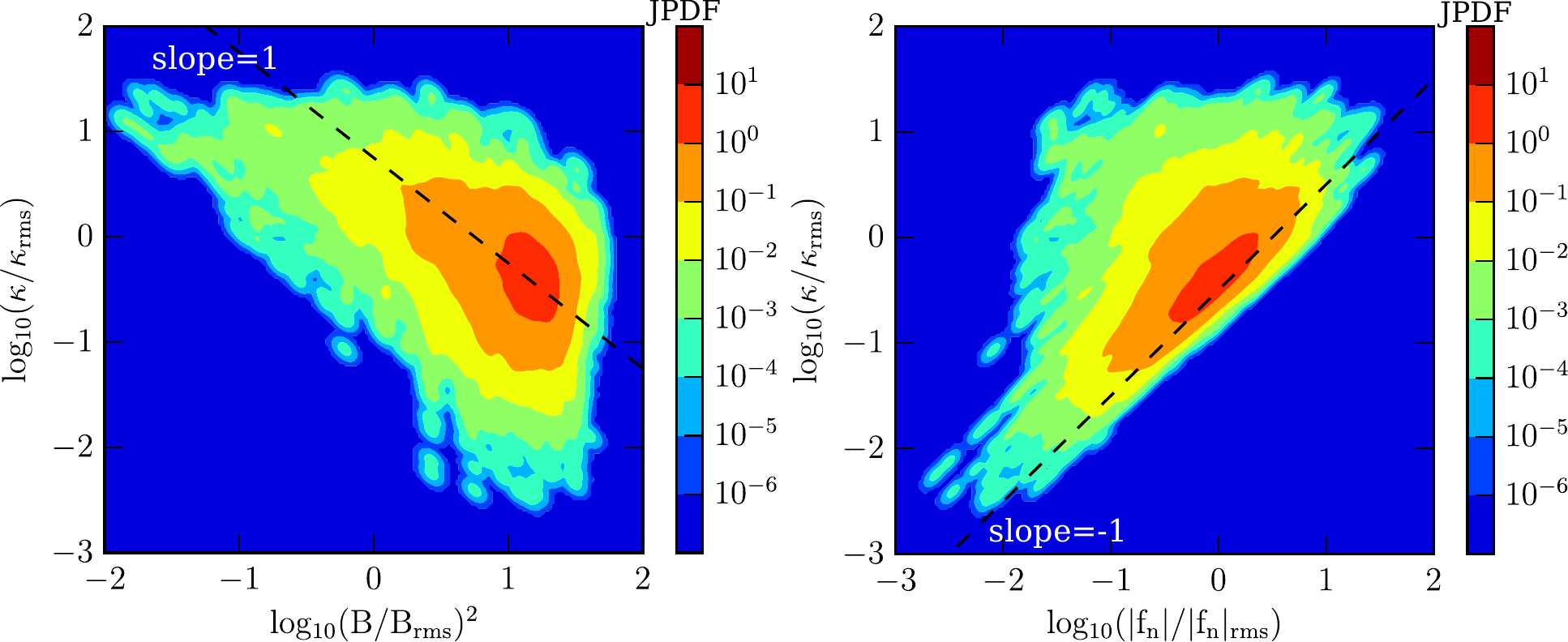}
		\caption{Joint PDFs of curvature $\kappa$ and (left panel): the square of magnetic field magnitude $B^2$, and (right panel): the magnitude of the force $|f_n|$ acting normal to the field lines. All quantities are normalized to their respective rms values. Dashed lines with slope of $1$ and $-1$ are shown for reference.}
		\label{fig:hist}
	\end{center}
\end{figure*}

The main quantitative observational result
of this study
is contained in Fig.~\ref{fig:curv_pdf},
which shows the 
probability distribution function
of the curvature for the
$\approx$ 40 minute magnetosheath interval of MMS data described in Table \ref{table1} 
and shown in Fig.~\ref{fig:overview}.
Many prior studies~\citep[e.g.,][]{Slavin2003JGR, Shen2003JGR, Shen2007JGR, Rong2011JGR, Sun2019GRL, Akhavan-Tafti2019GRL} have explored curvature of the magnetic field in the magnetosphere for individual events or collection of structures, but as far as we are aware, this is the 
first detailed analysis of \textit{statistics} of magnetic-field 
curvature in turbulent space plasmas using observational data.

The probability distribution function shown in figure~\ref{fig:curv_pdf}, exhibits 
two distinct powerlaw regimes:
at low values of the curvature field, its distribution
scales roughly as $\kappa^{+1}$, while at 
high curvature values the distribution behaves as $\kappa^{-2.5}$.
This is remarkably similar to the empirical and theoretical findings of~\cite{Yang2019ArXiv}.

To further clarify the 
statistics of the 
magnetic curvature, 
Fig.~\ref{fig:hist}
shows the joint probability distributions
of curvature and 
squared magnetic field magnitude, and
curvature and normal force.
The relationship between high curvature 
and regions of 
weak magnetic field is 
corroborated by the former. In Fig.~\ref{fig:curv_pdf}, the power-law regimes are separated at $\kappa/\kappa_{\mathrm{rms}} \lessapprox 0.1$ and $\kappa/\kappa_{\mathrm{rms}} \gtrapprox 1$. From the left panel of Fig.~\ref{fig:hist}, the curvature and magnetic field are rather well associated for $\kappa/\kappa_{\mathrm{rms}} \gtrapprox 1$ and the association begins to weaken at $\kappa/\kappa_{\mathrm{rms}} \lessapprox 0.1$. The Pearson-r coefficient between $\kappa/\kappa_{\mathrm{rms}}$ and $(B/B_{\mathrm{rms}})^{2}$ for $\kappa/\kappa_{\mathrm{rms}} \geq 0.9$ has a magnitude of 0.53 and it is 0.005 for $\kappa/\kappa_{\mathrm{rms}} \leq 0.15$. This is consistent with the intuition 
that weak magnetic fields are easier to bend, and 
leads to the above-described
$\kappa^{-2.5}$ curvature 
distribution in the weak magnetic field
regime.

Similarly, the 
positive correlation between 
curvature and 
normal force per unit volume at low curvature is in Fig.~\ref{fig:hist}, right panel. Again, at small curvature values, $\kappa$ and $f_n$ are well correlated with a Pearson-r coefficient value of 0.12 for $\kappa/\kappa_{\mathrm{rms}} \leq 0.15$, However, at high-curvature value the association is not so clear, resulting in a Pearson-r value of 0.02 for $\kappa/\kappa_{\mathrm{rms}} \geq 0.9$. This supports the reasoning that leads to the 
$\kappa^{+1}$ behavior of the 
$\kappa$ distribution at the small values of curvature. These quantifications are summarized in Table~\ref{tab:corr}. 

\begin{table}
\caption{Pearson's-r coefficient between curvature and magnetic field strength, and normal force for low and high curvature ranges.}
\label{tab:corr}
\centering
  \begin{tabular}{lc|c|c}
    &\multicolumn{1}{c}{} & \multicolumn{2}{c}{\text{Range}} \\
    & & $\kappa/\kappa_{\mathrm{rms}} \leq 0.15$ & $\kappa/\kappa_{\mathrm{rms}} \geq 0.9$ \\
    \cline{2-4}
    & $\kappa/\kappa_{\mathrm{rms}}, (B/B_{\mathrm{rms}})^{2}$ & $r=0.005$ & $r=-0.53$ \\
    \cline{2-4}
    \smash{\rotatebox[origin=c]{90}{\text{Variables}}} & $\kappa/\kappa_{\mathrm{rms}}, |f_n| /|f_n|_{\mathrm{rms}}$ & $r=0.12$ & $r=0.02$ \\
  \end{tabular}
\end{table}

The interval shown in Fig.~\ref{fig:overview} is selected for no special reason other than its long duration and the  preliminary observation that it exhibits well-developed turbulence properties~\citep{Parashar2018PRL}. Same analyses on other turbulent magnetosheath intervals produce similar results (see Appendix~\ref{sec:add} and Fig.~\ref{fig:qprep}).

\begin{figure}
	\begin{center}
		\includegraphics[width=\linewidth]{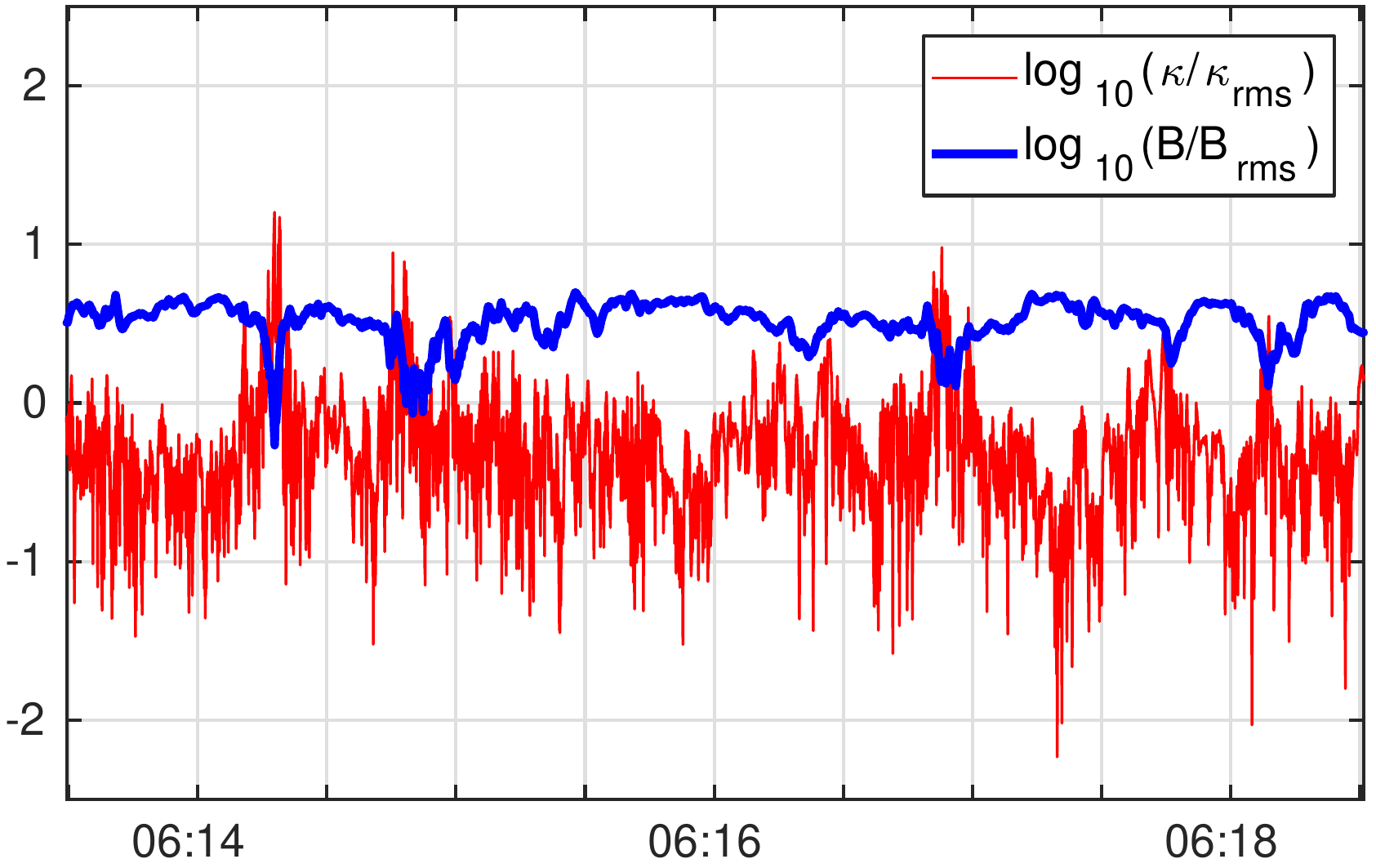}
		\includegraphics[width=\linewidth]{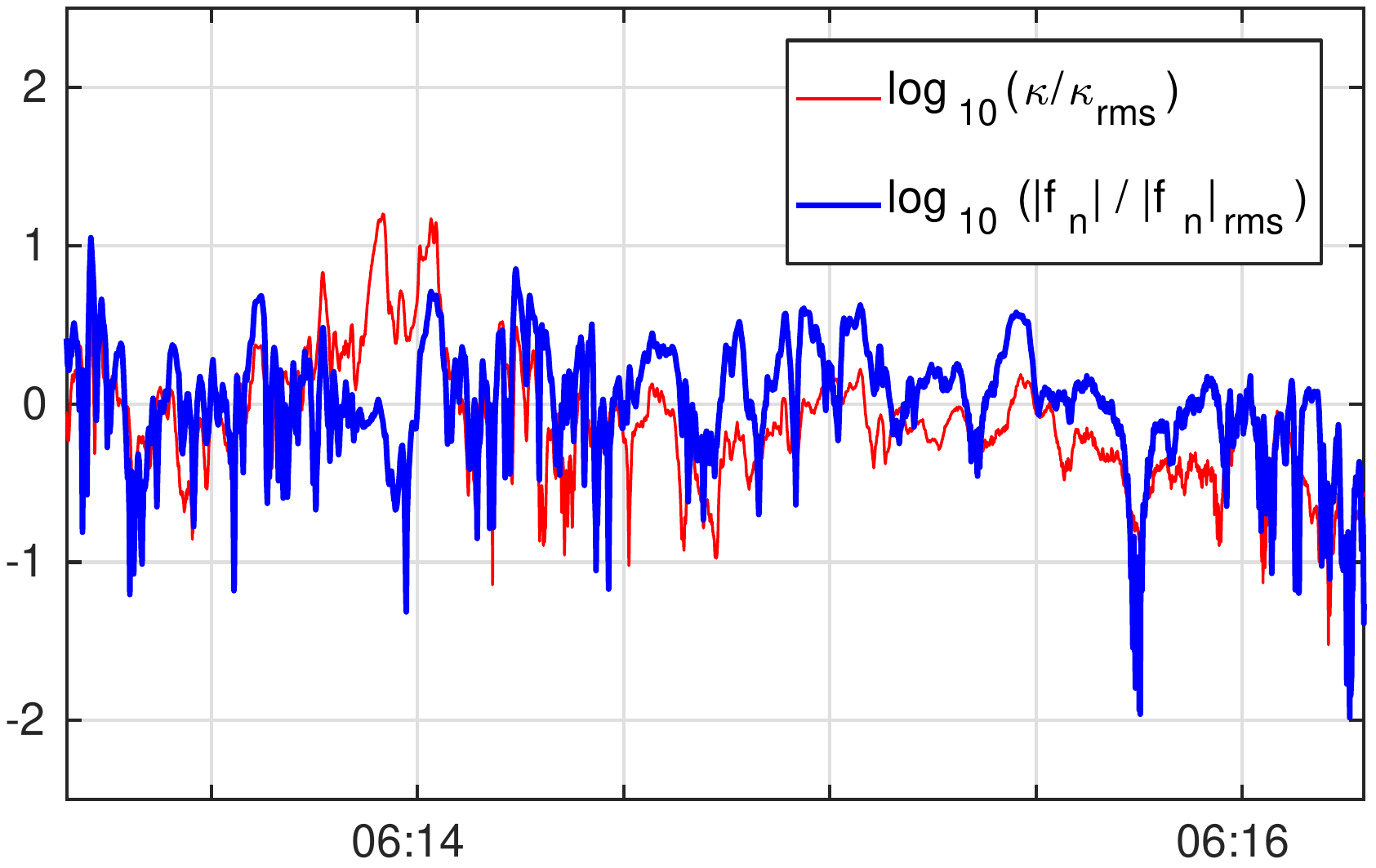}	
		\caption{Time series of the curvature field, in thin red line, superposed on the magnetic-field magnitude (Top panel) and the magnitude of the normal force (Bottom panel), in broad blue line, and for a small subinterval of the whole magnetosheath interval.}
		\label{fig:anticorr}
	\end{center}
\end{figure}

As a final, direct observational diagnostic,
in the two panels in Fig.~\ref{fig:anticorr}
we show small samples of
the time series of curvature and magnetic field (top panel), and curvature and the normal force (bottom panel). 
to illustrates how large (small) curvature regions
are often localized in regions of low  magnetic field strength (low normal force strength).

Indeed,
from the top panel in Fig.~\ref{fig:anticorr} 
one observes
several \textit{peaks} in curvature values 
that are contemporaneous
with sharp drops in the magnetic field strength. Similarly, from the bottom panel, strong \textit{dips} in the curvature is accompanied by dips in the normal force values.
For clarity, only small subintervals of the whole magnetosheath intervals are shown. We note that only curvature large enough, say $\kappa/\kappa_{\mathrm{rms}} \gtrapprox 1$ (i.e., the right power-law regime in Fig.~\ref{fig:curv_pdf}), is associated with small $B$, and only curvature small enough, say $\kappa/\kappa_{\mathrm{rms}} \lessapprox 0.1$ (i.e., the left power-law regime in Fig.~\ref{fig:curv_pdf}), is associated with small $f_n$. This behavior is also consistent with the trends seen in Fig.~\ref{fig:hist} and Table~\ref{tab:corr}.

\section{Discussion} \label{sec:disc}
Employing the unique capabilities of the \textit{MMS} mission,
we have studied the statistical properties of the curvature
of the magnetic field measured in the terrestrial magnetosheath by the FGM instrument onboard
each of the four spacecraft. The dataset employed is a long, 40 minute duration, 
burst mode interval in the terrestrial magnetosheath.
This determination
of the statistical character
of the magnetic curvature is the first of its kind 
in
a space plasma, as 
far as we are aware.

We find two powerlaw regimes in the distribution
of values of curvature: a $\kappa^{+1}$ regime
at low $\kappa$, and a
$\kappa^{-2.5}$ regime at large $\kappa$.
We also find 
an anticorrelation of curvature and magnetic field strength 
at low magnetic field strength, and 
a positive correlation of curvature and 
normal force per unit volume at small values of the 
force.
All of these results are consistent with
the findings of \cite{Yang2019ArXiv},
based on MHD simulations.
What is more remarkable is the degree
of quantitative agreement
of the present observations with the MHD results.
The simple theory outlined here,
clearly is adequate to 
explain the two powerlaw
ranges in the 
curvature that are seen in both simulations
and observations. 

It is interesting to note that the kind of distribution we find for curvature (Fig.~\ref{fig:curv_pdf}) has been studied in applied mathematics and is known as a ``double Pareto" distributions~\citep{REED2001EL, Reed2002PRE, Reed2004CSTM, Fang2012Book}, generalizing the standard nomenclature of Pareto distribution for a range of scale invariant power law behavior~\citep{Mitzenmacher2004IM}. This type of distribution generally indicates a multiplicative process. However, if such a process is uninhibited it leads to a log normal distributions. When a physical effect, such as the inner scale of turbulence, or the particle gyro motion changes the physics and limits the process, it becomes a double-Pareto. The mathematics of such processes may provide fruitful directions for additional study of the nature of magnetic field curvature and its effects on particle acceleration.

One possibility that 
presents itself 
is that these 
results 
may be applicable 
to turbulent 
magnetic fields
in other 
venues including 
other 
heliospheric environments 
and perhaps in
astrophysical contexts as well. We note that, in order to derive the power-law scalings, equations~(\ref{eq:Pklow}) and~(\ref{eq:Pk-high}), we assume that the magnetic field components are isotropically distributed. Real systems are never perfectly isotropic at any length scale, but the magnetosheath conditions are rather close to isotropy with a weak DC field (see figure~\ref{fig:overview} and table~\ref{tab:overview}). Extending the present study to other plasma systems, e.g., solar wind, magnetotail, magnetosphere, etc., would require appropriate modification to the derivation, although the basic arguments are expected to remain unchanged. Independent 
confirmation 
from other simulations, laboratory experiments, as well
as other observations, if available, 
is called for.
To the extent that 
these results are robust,
at least one major theoretical application is 
suggested.
Specifically,
curvature drift acceleration
theory~\citep[e.g.,][]{Hoshino2001JGR, Dahlin2014PoP, Guo2015ApJ},
has apparently been 
very successful 
in explaining electron energization in individual
magnetic reconnection events. 
Since this theory 
depends explicitly 
on $\kappa$,
one would 
expect that an immediate extension 
based on the present results
would be to include a statistical distribution of 
curvature values, to develop 
a curvature 
drift energization mechanism
appropriate to magnetized 
plasma turbulence.

\acknowledgments
This research was partially supported by NASA
under the Magnetospheric Multiscale Mission (\textit{MMS}) Theory and Modeling
program grant NNX14AC39G
and by NASA Heliospheric Supporting Research Grant
NNX17AB79G. We would like to
acknowledge the assistance of the \textit{MMS} instrument teams,
especially FPI and FIELDS, in preparing the data, as well as
the work done by the \textit{MMS} Science Data Center (SDC). The data used in this work are Level 2 FIELDS data products, in cooperation with the
instrument teams and in accordance with their guidelines. All \textit{MMS} data used in this study are publicly available at the \textit{MMS} Science Data Center (\url{https://lasp.colorado.edu/MMS/sdc/public/}). The Interplanetary Magnetic Field (IMF) data, measured by the  \textit{Wind} spacecraft, are used to determine the angle between the shock normal and the IMF in Appendix~\ref{sec:gaussian}. The IMF, shifted to Earth's bow-shock nose, can be found at \url{https://omniweb.gsfc.nasa.gov/}. The authors thank the \textit{Wind} team for the \textit{Wind} magnetic field data. 

\appendix

\section{Distribution of Normal Force at Small Values}\label{sec:fndist}
\begin{figure}[hb!]
	\begin{center}
		\includegraphics[width=0.5\linewidth]{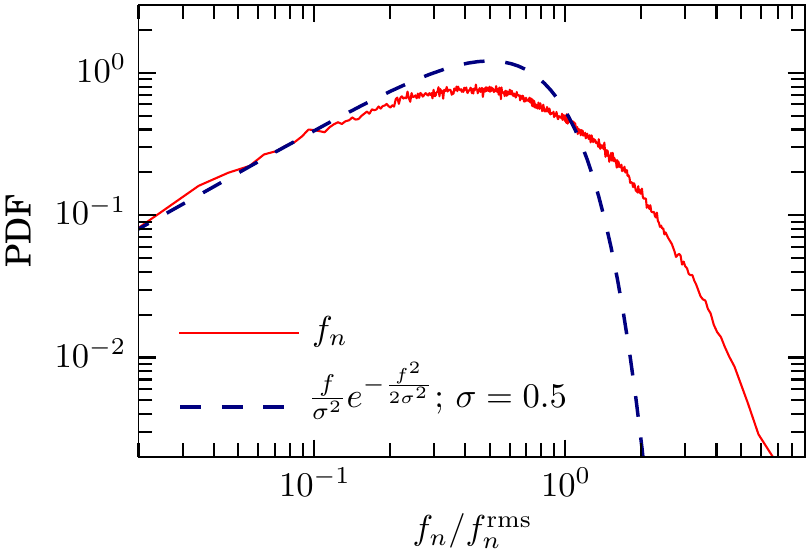}
		\caption{PDFs of the value of the normal force  $f_n$ as computed from \textit{MMS} data analyzed using a 40-minute interval.}
		\label{fig:fndist}
	\end{center}
\end{figure}
In deriving the scaling properties of curvature, $\kappa$, at low-$\kappa$ values (Eq.~\ref{eq:Pklow}), we assume that the normal force components $f_1$ and $f_2$ are independent Gaussian variables for small values.
However, the only result that is actually used in the subsequent development is Eq.~\ref{eq:pdf-fn} which is the distribution of the values of the \textit{magnitude} of the normal force, $f_n$. The exact form of the distributions of $f_1$ and $f_2$, therefore, is not a salient point. Rather, if $f_n$ follows a chi-squared distribution with 2 degrees of freedom (Eq.~\ref{eq:pdf-fn}) for small values of $f_n$, that would support the subsequent development of the theory. Fig.~\ref{fig:fndist} shows the distribution of $f_n$ and compares it with Eq.~\ref{eq:pdf-fn} for the interval analyzed in the main text. It is evident that the small values of normal force are well described by Eq.~\ref{eq:pdf-fn}.

\section{Distribution of Magnetic-Field Components in the Magnetosheath Plasma}\label{sec:gaussian}
In deriving the scaling properties of curvature, $\kappa$, at high-$\kappa$ values (Eq.~\ref{eq:Pk-high}), we assume that the probability distribution of magnetic-field components is approximately Gaussian, e.g., in Eq.~\ref{eq:Bxyz}. Although established in the pristine solar wind at 1 au~\citep{PadhyeEA01}, and expected in general for primitive variables in turbulence~\citep{Batchelor51}, the Gaussianity of the turbulent fluctuations in the magnetosheath has not been previously quantified, as far as we are aware~\citep[although see][]{Whang77}. To justify this approximation, here we examine the probability distribution functions (PDFs) of the  fluctuations of the magnetic field components using analysis of several \textit{MMS} data sets.
Functional fits as well as moment comparisons (kurtoses) are used in drawing conclusions concerning the degree of non-Gaussianity. 

\begin{figure}[hb!]
	\begin{center}
		\includegraphics[width=\linewidth]{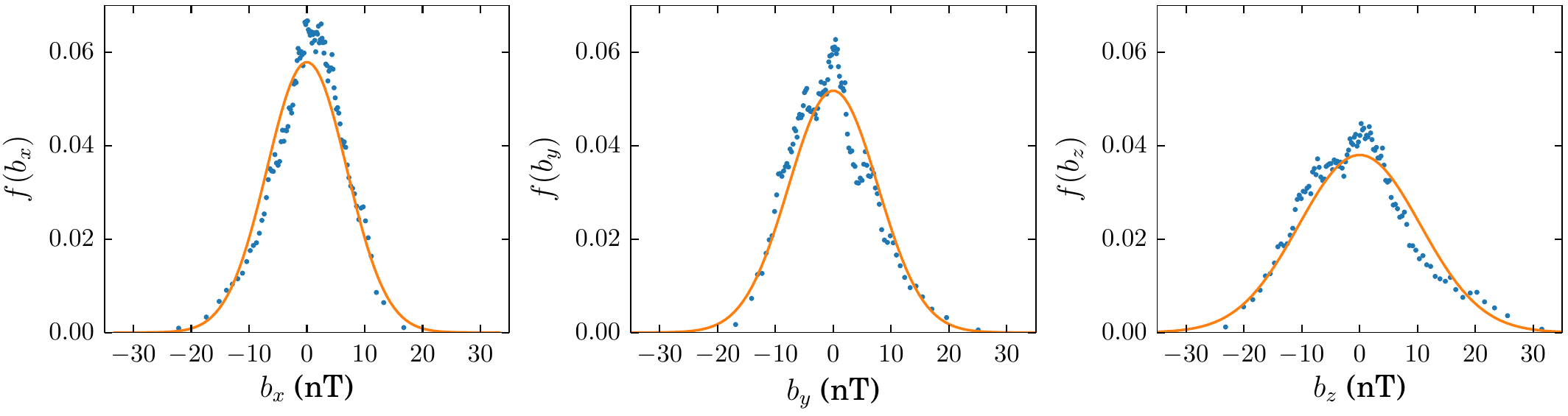}
		\caption{PDFs of fluctuations of the magnetic field as computed from \textit{MMS} data analyzed using a 40-minute interval. X, Y, and Z components are shown in top, middle, and bottom panel. The dots  represent centers of the binned data, and the solid line shows the reference Gaussian. Each of the 100 bins contains 15373 measurements.}
		\label{fig:gaussian}
	\end{center}
\end{figure}

To begin, we analyze the main 
40 minute data interval analyzed in the text, 
shown in Fig. \ref{fig:overview} and described in Table \ref{table1}.   The data from each spacecraft are rank-ordered 
into 100 bins of variable width such that each bin has an equal number of data points. For each component, data from all 4 spacecraft are collected together to increase the statistical weight. The density of points in each bin provides an approximation to the PDFs,
This procedure is carried out for 
the fluctuating 
magnetic-field components, and these empirically determined PDFs are shown in Fig.~\ref{fig:gaussian}. The solid curves are the corresponding Gaussian PDFs 
with zero mean and a variance equal to that computed from the data. The goodness of fit is measured by $\chi^{2}$, defined as
\begin{eqnarray}
\chi^{2} = \frac{\sum_{i} {\left[f(a_{i}) - g(a_{i}) \right]}^{2}\,\Delta a_{i} }{\sum_{i} {\left[f(a_{i}) \right]}^{2}\,\Delta a_{i}} \label{eq:chi},
\end{eqnarray}
where $f(a_{i})$ is the observed PDF of any magnetic-field component, and $g(a_{i})$ is the corresponding Gaussian distribution. For a perfect agreement $(f=g)$, $\chi^{2}=0$; small value of $\chi^{2}$ indicates satisfactory fitness. 

Quantitative results for 
the main 40 minute MMS interval are shown 
in the first row entries of Table~\ref{tab:kurtosis}.
As figures of merit, 
the values of the parameter $\chi^{2}$ are
listed along with values of the kurtosis
for each component. 
\begin{table}
	\caption{Kurtoses and $\chi^{2}$ values for the magnetic field components measured by \textit{\textit{MMS}} in the magnetosheath.} 
	\centering
	\label{tab:kurtosis}
	\begin{tabular}{l c c c c c c c}
		\hline 
		 \hspace{2cm} Interval & MMS  position & Parameter & $b_x$ & $b_y$ & $b_z$ & $B_{\mathrm{rms}} /|\langle \mathbf{B} \rangle|$ & Shock Type \\
		 & (X, Y)$_{\mathrm{GSE}}$ & & & & & &\\
		\hline
		26 December 2017,\hfill 06:12:43 - 06:52:23 & $(10\, \mathrm{R_e}, 9\, \mathrm{R_e})$ & $\chi^{2}$ & $0.018$ & $0.023$ & $0.023$ & 0.8 & quasi-$\parallel$ \\ & & Kurtosis & $3.56$ & $3.50$ & $3.23$ & & \\
		11 January 2016,\hfill 00:57:04 - 01:00:34 & $(9\, \mathrm{R_e}, -8\, \mathrm{R_e})$ & $\chi^{2}$ & $0.031$ & $0.020$ & $0.067$ & 1.5 & quasi-$\perp$ \\ & & Kurtosis & $3.76$ & $3.06$ & $4.5$ & & \\
	18 January 2017,\hfill 00:45:54 - 00:49:42 & $(8\, \mathrm{R_e}, -5\, \mathrm{R_e})$ & $\chi^{2}$ & $0.020$ & $0.067$ & $0.113$& 1.8 & quasi-$\parallel$ \\ & & Kurtosis & $2.77$ & $2.43$ & $3.56$ &  & \\
	27 January 2017,\hfill 08:02:03 - 08:08:03 & $(11\, \mathrm{R_e}, 6\, \mathrm{R_e})$ & $\chi^{2}$ & $0.04$ & $0.09$ & $0.02$& 2.1 & quasi-$\parallel$ \\ & & Kurtosis & $3.14$ & $3.08$ & $3.29$ &  & \\	
	21 December 2017,\hfill 06:41:55 - 07:03:51 & $(13\, \mathrm{R_e}, -1\, \mathrm{R_e})$ & $\chi^{2}$ & $0.085$ & $0.022$ & $0.019$& 2.1 & quasi-$\parallel$  \\ & & Kurtosis & $3.23$ & $2.92$ & $2.75$ &  & \\	
	21 December 2017,\hfill 07:21:54 - 07:48:01 & $(14\, \mathrm{R_e}, 0\, \mathrm{R_e})$ & $\chi^{2}$ & $0.012$ & $0.094$ & $0.045$& 1.9 & quasi-$\parallel$  \\ & & Kurtosis & $4.03$ & $2.83$ & $2.61$ & & \\
	19 April 2018,\hfill 05:08:04 - 05:41:51 & $(-3\, \mathrm{R_e}, -22\, \mathrm{R_e})$ & $\chi^{2}$ & $0.011$ & $0.014$ & $0.011$& 3.1 & quasi-$\perp$  \\ & & Kurtosis & $3.41$ & $3.81$ & $3.46$ &   &  \\
	23 April 2018,\hfill 07:50:14 - 08:33:41 & $(3\, \mathrm{R_e}, 18\, \mathrm{R_e})$ & $\chi^{2}$ & $0.019$ & $0.035$ & $0.027$& 1 & quasi-$\perp$  \\ & & Kurtosis & $3.54$ & $3.47$ & $3.73$ &  & \\
	27 October 2018,\hfill 09:13:14 - 09:57:41 & $(-2\, \mathrm{R_e}, 24\, \mathrm{R_e})$ & $\chi^{2}$ & $0.017$ & $0.010$ & $0.029$ & 2.5 & quasi-$\parallel$  \\ & & Kurtosis & $3.07$ & $3.28$ & $2.86$ &  & \\
	21 November 2018,\hfill 16:10:14 - 16:55:31 & $(11\, \mathrm{R_e}, 13\, \mathrm{R_e})$ & $\chi^{2}$ & $0.010$ & $0.049$ & $0.009$ & 0.9 & quasi-$\perp$  \\ & & Kurtosis & $3.84$ & $3.75$ & $2.92$ &  & \\
	29 November 2018,\hfill 22:42:34 - 23:31:01 & $(11\, \mathrm{R_e}, 8\, \mathrm{R_e})$ & $\chi^{2}$ & $0.008$ & $0.005$ & $0.008$ & 5 & quasi-$\parallel$  \\ & & Kurtosis & $3.04$ & $2.90$ & $3.10$ &  & \\
	05 December 2018,\hfill 14:53:23 - 15:20:13 & $(12\, \mathrm{R_e}, 7\, \mathrm{R_e})$ & $\chi^{2}$ & $0.019$ & $0.015$ & $0.013$ & 7.5 & quasi-$\parallel$  \\ & & Kurtosis & $3.39$ & $2.95$ & $2.82$ &  & \\
	11 January 2019,\hfill 03:22:23 - 03:52:23 & $(12\, \mathrm{R_e}, 2\, \mathrm{R_e})$ & $\chi^{2}$ & $0.011$ & $0.02$ & $0.028$ & 2.0 & quasi-$\parallel$  \\ & & Kurtosis & $3.52$ & $2.56$ & $2.94$ &  & \\	
	05 April 2019,\hfill 10:58:33 - 11:25:52 & $(12\, \mathrm{R_e}, -10\, \mathrm{R_e})$ & $\chi^{2}$ & $0.024$ & $0.035$ & $0.086$ & 1.9 & quasi-$\parallel$  \\ & & Kurtosis & $2.66$ & $2.69$ & $2.63$ &  & \\		
	\hline
	\end{tabular}
\end{table}
While the goodness of the Gaussian 
representation is measured by $\chi^{2}$,
the closeness of the PDFs to Gaussian distributions can be also be quantified by the kurtosis. The kurtosis for a Gaussian distribution is 3;  a kurtosis value greater (less) than 3 represents a super (sub)-Gaussian distribution~\citep{Matthaeus2015PTRSA}. The closeness of the PDFs of the magnetic-field components to Gaussian is clear from Fig.~\ref{fig:gaussian} and from the values of $\chi^2$ and kurtoses listed in Table~\ref{tab:kurtosis}. On deriving the power law of the curvature PDF at large values, we also neglect the differences among $B_x$, $B_y$, $B_z$ distributions, i.e., the variances of the three components are the same in Eq.~\ref{eq:Bxyz}. For the 40-minute MMS interval we have, $\sigma_{B_{x}}=7,\,\sigma_{B_{y}}=8,\,\sigma_{B_{x}}=10$, which can also be seen from the the width of the three distributions shown in Fig.~\ref{fig:gaussian}.

To draw a more proper conclusion, we further analyze a few other turbulent magnetosheath intervals. Note that for good statistical weight, long intervals are required. Therefore, we select magnetosheath intervals of at least few minutes duration and with no prominent discontinuity. All the selected intervals have large fluctuation amplitude with $B_{\mathrm{rms}} /|\langle \mathbf{B} \rangle| \gtrsim 1$, and the variances of the three components close. Further, we check that each of these intervals exhibits a Kolmogorov ``$-5/3$" spectrum, which is often considered an adequate indicator of well-developed turbulence. The collection of studied magnetoshetah intervals are reported in Table~\ref{tab:kurtosis}, where we also report whether each interval is downstream of a quasi-parallel or quasi-perpendicular shock region. The nature of turbulent fluctuations may be significantly different in the magnetosheath plasma downstream of a quasi-parallel and a quasi-perpendicular shock. The plasma downstream of a quasi-parallel shock is usually found to be more turbulent, relative to that of a quasi-perpendicular shock. 
The examined intervals include those corresponding to both kinds of shock and with a substantial variation in the normalized fluctuation  amplitude $B_{\mathrm{rms}}/|\langle {\bf B}\rangle|$.

From Table~\ref{tab:kurtosis}, we see that the Kurtosis lies generally between 2.4 and 4.03, and the values of Kurtosis and the fitness parameter $\chi^2$ do not appear to change systematically from quasi-parallel to quasi-perpendicular shock. Although this is not a fully exhaustive sampling, it appears that turbulent magnetosheath fluctuations are often found in a near-Gaussian state, as is common for fluctuations of the primitive variables in strong homogeneous turbulence~\citep{Batchelor51,Schumann1978JFM}. 

\section{Additional Supporting Analysis}\label{sec:add}
\begin{figure}
	\begin{center}
		\includegraphics[width=\linewidth]{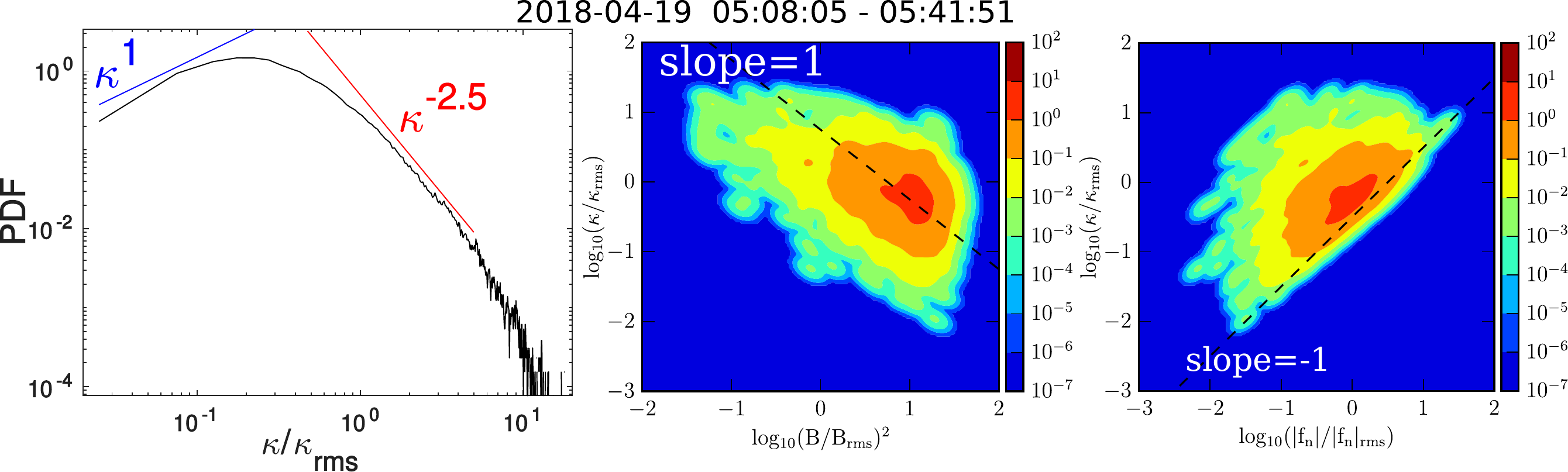}
		\caption{Magnetosheath plasma downstream of a quasi-perpendicular shock observed by MMS on 19 April 2018 (Table~\ref{tab:kurtosis}). Left: PDF of the magnetic field curvature $\kappa$ normalized to its rms value $\kappa_{\mathrm{rms}}$  Middle: Joint PDF of curvature $\kappa$ and the square of magnetic field magnitude $B^2$; Right: Joint PDF of curvature $\kappa$ and the magnitude of the force $|f_n|$ acting normal to the field lines.}
		\label{fig:qprep}
	\end{center}
\end{figure}
We also perform the analyses presented in the main article, namely Fig.~\ref{fig:curv_pdf} and Fig.~\ref{fig:hist}, for all the intervals listed in Table~\ref{tab:kurtosis}. Again, every interval is found to return reasonably similar result (not shown here), with no systematic variation between quasi-parallel and quasi-perpendicular shocked plasma. The interval studied in the main article corresponds to a quasi-parallel shock. For demonstration, we show the corresponding figures produced for a long magneotsheath interval downstream of a quasi-perpendicular shock, in Fig.~\ref{fig:qprep}. Again, the agreement is satisfactory, and the scaling laws appear to hold, in general, for turbulent magnetosheath plasmas.

\bibliographystyle{aasjournal}

\end{document}